\documentclass[prl,twocolumn,floatfix,notitlepage,nobibnotes,nofootinbib,showpacs,showkeywords]{revtex4}
\usepackage{amsmath}
\usepackage{amssymb}
\usepackage{bm}
\usepackage{graphicx}
\usepackage{color}
\usepackage [cp1251]{inputenc}
\usepackage [english]{babel}

\usepackage{hyperref}

\usepackage{graphicx} 
\usepackage{ulem}

\def\be{\begin{equation}}  
\def\ee{\end{equation}}  
\def\ba{\begin{eqnarray}}  
\def\ea{\end{eqnarray}}  
\def\bc{\begin{center}}  
\def\ec{\end{center}}

\begin{document}

\title{
Probing of electromagnetic fields on atomic scale by \\  photoelectric phenomena in graphene
}
\author{
P.~Olbrich,$^1$ C. Drexler,$^1$ L.E.~Golub,$^2$ S.~N.~Danilov,$^1$  V.~A.~Shalygin,$^3$ 
R.~Yakimova,$^4$ S.~Lara-Avila,$^5$ S.~Kubatkin,$^5$
B. Redlich$^6$, R. Huber,$^1$ S.~D.~Ganichev$^{1}$
}

\affiliation{$^1$ Terahertz Center, University of Regensburg,
93040 Regensburg, Germany}
\affiliation{$^2$ Ioffe Physical-Technical Institute, Russian Academy of Sciences,
194021 St.~Petersburg, Russia}
\affiliation{$^3$St.\,Petersburg State Polytechnic University, 195251 St.\,Petersburg, Russia}
\affiliation{$^4$
Link{\"o}ping University,
S-58183 Link{\"o}ping, Sweden}
\affiliation{$^5$
Chalmers University of Technology,
S-41296 G{\"o}teborg, Sweden}

\affiliation{$^6$ FOM Institute for Plasma Physics ``Rijnhuizen'', P.O.
Box 1207, NL-3430 BE Nieuwegein, The Netherlands}

\begin{abstract}
We report on the observation of the reststrahl band assisted 
photocurrents in epitaxial graphene on SiC excited 
by infrared radiation. The peculiar spectral dependence 
for frequencies lying within the reststrahl band of the 
SiC substrate provides a direct and noninvasive way to probe  the electric field magnitude
at atomic distances from the material's surface.
Furthermore our results reveal that nonlinear optical 
and optoelectronic phenomena in 2D crystals and other atomic scale 
structures can be giantly enhanced by a proper combination of the 
spectral range and substrate material. 
%
%
%
%
%
%
%
%
%
%
%
\end{abstract}

\date{\today}

\maketitle


Since the discovery of graphene, optical and optoelectronic properties of two dimensional (2D) 
crystals have attracted continuously growing attention~\cite{Novoselov2012}. 
A keen interest has been motivated by the prospective application of mono- or few-layer systems 
in nonlinear optics~\cite{Dean2009,Hendry2010,10.1063/1.3483872,Hotopan11,GlazovGanichev},
solar cells~\cite{Koppens2013},
displays~\cite{Bae2010}, optoelectronics~\cite{GuT.:2012fk},
sensors~\cite{Vicarelli2012,2011rangel},
or  plasmonic devices~\cite{Chen2012, Fei2012}.
 Bridging the size mismatch between macroscopic photonics and atomic-scale integrated electronics 
 all these concepts universally depend on a key quantity: the local optical fields acting on the charge carriers in 2D systems. 
These fields 
deviate from the emitted or incident  
waves due to the dielectric environment of the supporting substrate.
A mesoscopic  description has been invoked to obtain electromagnetic fields in the 
vicinity of a fictitious effective medium characterized by a mathematically sharp interface and 
bulk dielectric functions 
~\cite{Jablan2009, Koppens2011, Chen2012, Fei2012}. 
Yet, in the extreme limit, of an atomic thin system
a mesoscopic model is not justified a priori and electromagnetic fields may be altered by 
the modified polarization response of the surface structure.

At the same time, the precise knowledge of 
the local electromagnetic fields is particularly important for phenomena that scale nonlinearly 
with the field amplitude. Examples range from optical nonlinearities to high-frequency transport 
as studied in 2D crystals~\cite{Dean2009,Hendry2010,10.1063/1.3483872,Hotopan11,GlazovGanichev,Jablan2009,Koppens2011,Chen2012,Fei2012,karch2010,edge,PRB2010,Sun:2012ys,Prechtel:2012kx}, 
carbon nanotubes~\cite{cnt:book,ivch_spi,PhysRevA.63.053808,doi:10.1021/nl203003p}, topological insulators~\cite{Zhang2011,Hosur2011,McIver2012}
and single molecules~\cite{Lupton}. 
Nonetheless, measuring electric fields on atomic distances is challenging.
Surface-confined plasma oscillations in graphene have shown to change their dispersion 
sensitively in the spectral vicinity of the reststrahl band~\cite{Cardona_book} of 
substrates~\cite{Daas}. They could, 
thus, be basically used as sensors for the local dielectric environment. Yet,  
the depth resolution 
of the evanescent field remains on the order of several 100~nm. 
Even for near-field microscopy reaching extreme sub-wavelength spatial 
resolution~\cite{Fei2012} or a recent approach tracing the natural optical-frequency magnetic dipole transitions in lanthanide ions~\cite{Taminiau},
the atomic scale has been out of reach.

\begin{figure}[t]
\includegraphics[width=\linewidth]{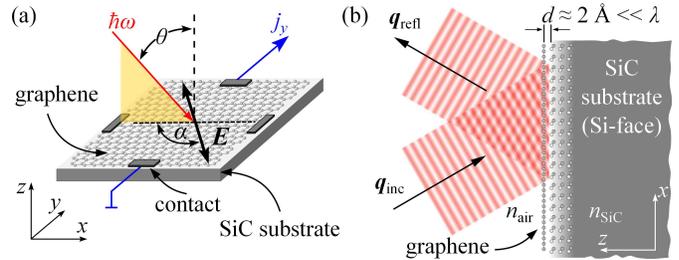}
\caption{(a) experimental geometry (b) reflection of a plane wave 
from a graphene layer on a SiC substrate deposited at the distances of about 2~\AA \, 
from the surface. 
} 
\label{fig00_geometry}
\end{figure}

Here, we demonstrate that measurements of high 
frequency photoelectric effects
in graphene deposited on a substrate 
provide a direct 
way to probe the electric field magnitude, $E_0$,  at atomic distances from the material's surface.
We show that second-order photoelectric effects excited in 
graphene exhibit a peculiar spectral dependence for frequencies lying 
within the reststrahl band of the SiC substrate.
The resonances of the photocurrents are
attributed to the variation of the out-of-plane and in-plane components of the 
radiation electric field acting on electrons that are confined in 
the graphene layer deposited at a distance $d \approx$ 2~\AA \, 
from the SiC surface, see Fig.~\ref{fig00_geometry}(b) and Ref.~\onlinecite{Borysiuk}.
We show that an analysis of the field distribution based on the macroscopic Fresnel formulas 
surprisingly well describes all experimental findings, 
while there remain quantitative discrepancies.
As an important result, the observed reststrahl band assisted photocurrent also
clearly demonstrates that nonlinear optical 
and optoelectronic phenomena in 2D crystals, carbon nanotubes and topological insulators 
can be giantly enhanced by a proper combination 
of the spectral range and substrate material.

\begin{figure}[t]
\includegraphics[width=0.8\linewidth]{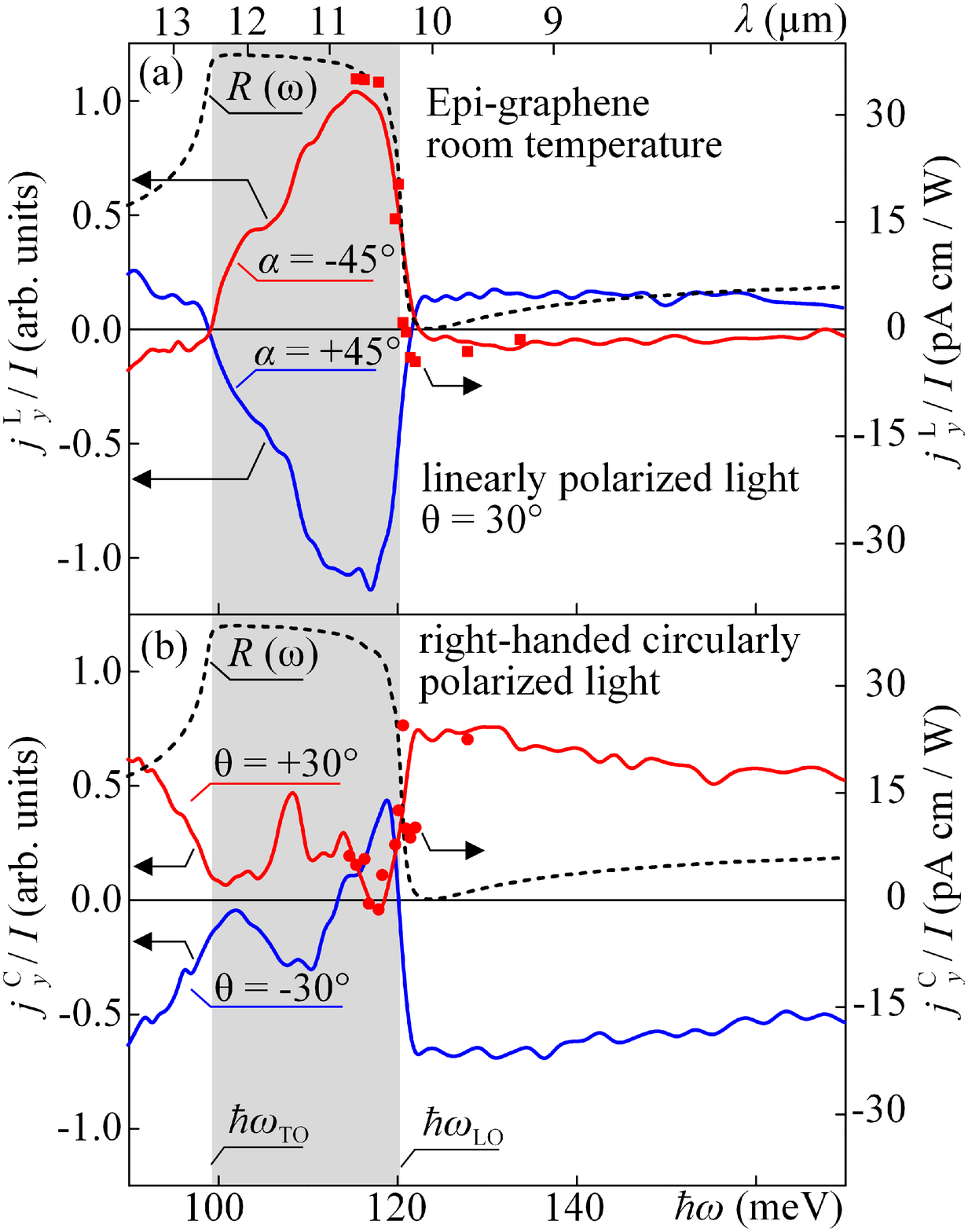}
\caption{Spectra of the photocurrent
excited by (a) linearly 
and (b) circularly  
polarized radiation. 
Solid lines and circles show the data obtained with free electron and cw CO$_2$ lasers, respectively.
Dashed lines show the reflection spectra $R(\omega)$, 
which reveals a clear reststrahl band of SiC (gray area) with maximum reflection of about 100\%.
} 
\label{fig01_spectr}
\end{figure}

Photocurrents have been observed in 
several large area \textit{n}-type graphene monolayer samples~\cite{edge,naturenano} at room temperature.
Figure~\ref{fig00_geometry}(a) depicts the corresponding 
experimental geometry. To induce the photocurrents we used the radiation of the pulsed free electron 
laser ``FELIX'' operating at tunable wavelength $\lambda$ from 7 to 16~$\mu$m 
or a \textit{cw} CO$_2$ laser. Radiation was applied in the ($xz$) plane at an 
angle of incidence $\theta$, varied between $-30^\circ$ to +30$^\circ$ to the 
layer normal, $z$. Details on the sample preparation and experiments are 
given in the supplementary material.
%

Illuminating an unbiased graphene layer with polarized radiation at oblique
incidence we detected a photocurrent signal 
whose spectral behaviour is shown in Fig.~\ref{fig01_spectr}.
Panel (a) shows the photocurrent $j_y^L$ excited by linearly polarized radiation 
with the azimuth angle $\alpha = \pm 45^\circ$ (see Fig.~\ref{fig00_geometry}(a)). 
A remarkable observation is that the current in the range of photon 
energies $\hbar \omega$ between about 99 and 120 meV changes sign while its value increases by more than an order of magnitude
compared to that excited by light with lower and higher frequencies.
%
 Despite the fact that the photocurrent shows a peculiar spectral 
dependence the overall functional behaviour of the photocurrent 
remains unchanged in the whole frequency range: The current
i)~ scales linearly with the radiation intensity $I \propto E_0^2$ 
ii)~is characterized by a short response  time and iii)
varies with the azimuthal angle $\alpha$ and the angle of incidence $\theta$ 
as $j_y^L = L   \sin 2\alpha \sin{\theta} E_0^2$, 
where $L$ is a prefactor. 
%
In particular, it reverses it's direction by changing  $\alpha$ 
from $+45^\circ$  to $-45^\circ$ for a fixed  $\theta$, 
see Fig.~\ref{fig01_spectr}(a) as well as upon switching 
$\theta = +30^\circ$ to $-30^\circ$ (not shown).
%
Studying the radiation reflection, see dashed lines in Fig.~\ref{fig01_spectr} and supplemantary materials,  
reveals that the position and width of the photocurrent resonance matches well the reststrahl band of the 
SiC~\cite{Patrick,Daas}, 
%
which is characterized by an almost total reflection $R(\omega)$ between the longitudinal, $\hbar \omega _{\rm {LO}}$,   
and transversal, $\hbar \omega _{\rm {TO}}$, optical phonon energies. 

\begin{figure}[t]
\includegraphics[width=\linewidth]{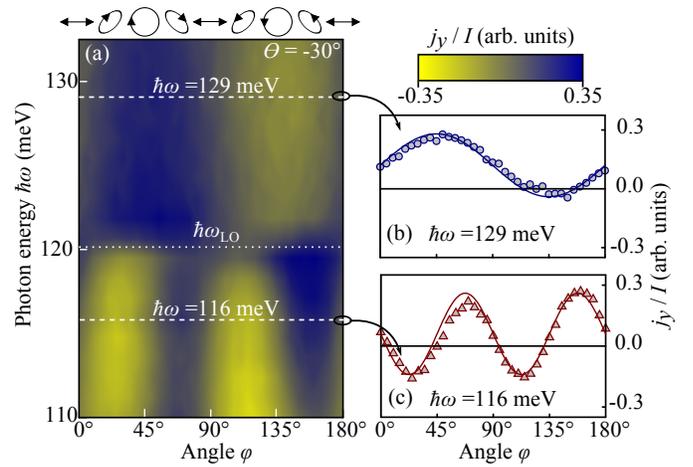}
\caption{
(a) Photocurrent $j_y/I$ as a function of the phase angle $\varphi$ defining 
radiation helicity and photon energy $\hbar \omega$. 
(b) and (c) show corresponding dependences obtained for 
 $\hbar\omega = 129$~meV 
and 116~meV, respectively.
The ellipses on top of the panel (a) illustrate the polarization states for 
several angles $\varphi$.
}
\label{fig02_phi_dep}
\end{figure}

%
%

The photocurrent is also observed for elliptically (circularly) polarized
radiation, obtained via rotation of Fresnel rhombus by the angle $\varphi$. 
By that we controllably vary the degree of linear, $P_{\rm l} = \sin 4 \varphi /2$, and circular, $P_{\rm c} = \sin 2 \varphi$, polarization, respectively.
%
%
Figure~\ref{fig02_phi_dep}(a) shows the polarization
dependence of $j_y$ measured for different photon energies in the
vicinity of the reststrahl band. The detected 
photocurrent 
can be well fitted by $j_y =  j^{L}_y +   j^{C}_y= [(L/2)  \sin 4\varphi  + C \sin 2\varphi] \sin{\theta} \, E_0^2$
as shown in Figure~\ref{fig02_phi_dep}(b) and (c) for photon energies below and above $\hbar \omega _{\rm {LO}}$, respectively. 
Here the linear photocurrent $j^{L}_y$, given by the fourth  harmonics of
$\varphi$, is just the above discussed current in response to linearly 
polarized radiation but excited by elliptically polarized light.
By contrast, the circular current $j^{C}_y$ 
stems from the radiation helicity and is proportional to the degree of circular polarization $P_{\rm c}$:
It achieves a maximum for circularly polarized radiation and 
reverses its direction by switching from right- to left-handed polarized light.
Figure~\ref{fig02_phi_dep} reveals 
the crossover from the dominating circular ($\propto \sin{2\varphi}$) to linear 
($\propto \sin{4\varphi}$) photocurrent contributions in the vicinity of $\hbar \omega _{\rm {LO}}$. 
The spectral behaviour of the circular contribution, $j^{C}$, is shown for right-handed circular polarization in 
Fig.~\ref{fig01_spectr} (b).
Similar to the linear photocurrent we found here a peculiar spectral behaviour within the reststrahl band.
However, by contrast to $j_y^L$, the circular photocurrent is not enhanced but rather suppressed.
It has a complex spectral behaviour exhibiting a change of sign and a dip close to the center of the reststrahl band.
Comparing circular and linear photocurrent spectra in the whole investigated range we see that 
the circular photocurrent $j^{C}_y$ dominates for frequencies  below $\hbar \omega _{\rm {TO}}$ and above $\hbar \omega _{\rm {LO}}$, 
while the linear current $j^{L}_y$ is almost vanishing outside of the reststrahl band.



The microscopic origin of the photocurrent under study outside of the reststrahl 
band has been previously investigated in Ref.~\cite{PRB2010}.
It has been demonstratet that the current is caused  by a sum of the photon
drag (PDE) and  photogalvanic (PGE) effects of comparable strength.
In particular, it has been shown that the photon drag effect in graphene is caused by a simultaneous action of the
electric and magnetic field components of the infrared radiation and, in fact, can be classified as a dynamic Hall effect~\cite{karch2010}.
The origin of the photogalvanic effect is the asymmetry of electron scattering induced by radiation and structure inversion asymmetry~\cite{PRB2010,Ganichev02}.
The addressed above fact that all 
characteristic features of the photocurrent
are the same within and outside the reststrahl band, 
indicates that its microscopic origin remains unchanged. 
Actually, this is not surprising because for normal incidence, $\hbar \omega \ll E_{\rm F}$,
and room temperature no resonances are expected for the light-matter interaction in pristine graphene.
However, one can expect dramatic modifications of local electric fields acting on carriers in 
graphene for frequencies within the reststrahl band of the substrate, which is characterized by a negative dielectric 
constant of the material. Indeed the coincidence of the increased reflection with the observed 
resonance of the photocurrent clearly indicates the common origin of both effects. 

As we show below, the resonant photoresponse excited in graphene by radiation with frequencies 
within the reststrahl band of the substrate can be well understood  considering only the spectral behaviour of the in-plane
and out-of-plane radiation electric field components, without going into microscopic details.
The required electric field components responsible for the photocurrent formation provides the phenomenological theory of PDE and PGE.
In line with the experiment we consider the transverse photocurrent 
$j_y$ generated in the direction perpendicular to
the incidence plane $(xz)$. 
Following Ref.~\onlinecite{GlazovGanichev} the current density due to the photon drag effect is given by
\begin{equation}
\label{j_PDE}
    j_y^{PDE} = T q_x  (E_x E_y^* + E_x^* E_y) + T' q_x  {\rm i} (E_y E_x^* - E_x E_y^*),
\end{equation}
and due to the photogalvanic effect by
\begin{equation}
\label{j}
j_y^{PGE} = \chi \, (E_z E_y^* + E_y {E_z}^*)
+ \gamma \, {\rm i} (E_z E_y^* - E_y {E_z}^*).
\end{equation}
Here $E$ is the electric field acting on electrons, $T$ and $\chi$   are coefficients describing, respectively, the 
PDE and PGE~\cite{footnote2} currents proportional to the linear polarization degree $P_l$ given by symmetrical combinations of electric field components.
 The two remaining coefficients correspond to the circular PDE ($T'$) and PGE ($\gamma$) currents. 
These contributions reverse the direction upon switching the photon helicity $P_{\rm c} = i(\bm E \times \bm E^*)\cdot {\bm q}/q$.

In the following analysis we assume coefficients $\chi$, $\gamma$, $T$ and $T'$ 
to be frequency independent in the narrow frequency range of the reststrahl band  
and focus on the frequency behaviour of the electric field components only.
This assumption is reasonable for the considered experimental conditions because
for room temperature, $\hbar\omega \ll E_{\rm F}$, and $\omega\tau \sim 1$,  
the radiation absorption is caused by Drude-like indirect intraband optical transitions. 
Hence,  $\chi$, $\gamma$, $T$ and $T'$ have smooth frequency dependences~\cite{karch2010,PRB2010}. 
Moreover, we disregard a possible influence of graphene itself on the electric 
field magnitudes.

To obtain the frequency dependence of the required electric field components 
we use macroscopic Fresnel formulas, which, strictly speaking, are applicable for 
representation of dielectric medium by a homogeneous function $\varepsilon(\omega)$, which is independent of the position within the medium.
In this approach the electric fields are formed by superposition 
of the incident and reflected waves, see Fig.~\ref{fig00_geometry}(b). 
They are described by the corresponding Fresnel transmission coefficients, and, consequently,
reflect the spectral behaviour of the dielectric function of the substrate, $\varepsilon(\omega)$. 
As it is well known, the latter
exhibits a strong anomaly within the reststrahl band~\cite{Cardona_book} 
\begin{equation}
\label{eps}
\varepsilon(\omega) = \varepsilon_\infty + {\varepsilon_0-\varepsilon_\infty \over 1 - (\omega/\omega_{TO})^2 - {\rm i}\omega\Gamma/\omega_{TO}^2},
\end{equation}
where $\omega_{TO}$ and $\Gamma$ are the frequency and the damping
of the TO phonon, $\varepsilon_0$ and $\varepsilon_\infty$ 
are the low- and high-frequency dielectric constants, respectively. 
These dielectric constants relate $\omega_{TO}$ with the LO phonon frequency 
via the Lyddane-Sachs-Teller relationship, $\omega_{LO}=\sqrt{\varepsilon_0/\varepsilon_\infty}\omega_{TO}$.
The complex dielectric function and, consequently, the complex 
refractive index $\sqrt{\varepsilon}=n+{\rm i}\varkappa$, 
determine the frequency dependence of the electric field components.
Here $n$  is refractive index and $\varkappa$ the extinction coefficient, see supplementary materials.
The in-plane components $E_x$ and $E_y$ 
are continuous and found from the Maxwell equation ${\rm div} \bm E =0$ yielding: 
\begin{equation}
\label{E_xy}
    E_x = t_p E_{0p} (n + {\rm i}\varkappa) \Xi,  
    \qquad
    E_y = t_s E_{0s},
\end{equation}
where $E_{0s}$, $E_{0p}$ are the corresponding parts of the incident wave amplitude, $\Xi = 1 / \sqrt{n^2+\varkappa^2+\sin^2{\theta_0}}$,
and $t_{s}$, $t_p$ are the standard Fresnel amplitude transmission coefficients for 
$s$- and $p$-polarizations,  see supplementary materials.

While the in-plane field components are continuous at the air/SiC interface, 
its normal component $E_z$ is discontinuous having different values inside and 
outside the substrate given by:
\begin{equation}
\label{Ez}
E_z^{in} = -t_p E_{0p} \sin{\theta_0} \Xi,
\quad   E_z^{out} = \varepsilon E_z^{in}.
\end{equation}
%

Substituting the above components of the radiation electric field in Eqs.~(\ref{j_PDE}) and (\ref{j}) 
we obtain  spectral behaviour of the photon drag and photogalvanic currents, respectively. 
%
%
%
We start with the PDE current given by the in-plane components $E_{x,y}$ and  
photon wavevector $q_x=(\omega/c)\sin{\theta_0}$. 
All these quantities are continuous, therefore, 
%
from Eqs.~\eqref{j_PDE} and~\eqref{E_xy} we obtain one solution for each, linear and circular, photocurrents:
%
\begin{align}
    j_y^{L} =&  {\omega\over c}\sin{\theta_0} P_{\rm l} |E_0|^2   \\
		& \times \left[ (nT + \varkappa T')\,{\rm Re}(t_p^* t_s) + (\varkappa T - nT') \, {\rm Im}(t_p^* t_s) \right] \Xi, \nonumber \\
        j_y^{C} =&  {\omega\over c}\sin{\theta_0} P_{\rm c} |E_0|^2   \\ 
				& \times \left[(\varkappa T - nT')\,{\rm Re}(t_p^* t_s) - (nT+\varkappa T') \, {\rm Im}(t_p^* t_s) \right] \Xi \nonumber.
\end{align}
%
The above equations yield at first glance a surprising result.
By contrast to Eqs.~(\ref{j_PDE}) the obtained linear and circular PDE currents are determined by both $T$ and $T'$ coefficients.
This comes from the fact that, within the reststrahl band, the radiation acting on the electrons in the graphene layer 
becomes elliptically polarized even for irradiation with purely linear or circular light~\cite{footnote3}.
Outside the reststrahl band $\varkappa$ and ${\rm Im}(t_p^* t_s)$ are almost 
zero. Thus the linear and circular PDE currents are given solely 
by the coefficients $T$ and $T'$, respectively.

\begin{figure}[t]
\includegraphics[width=\linewidth]{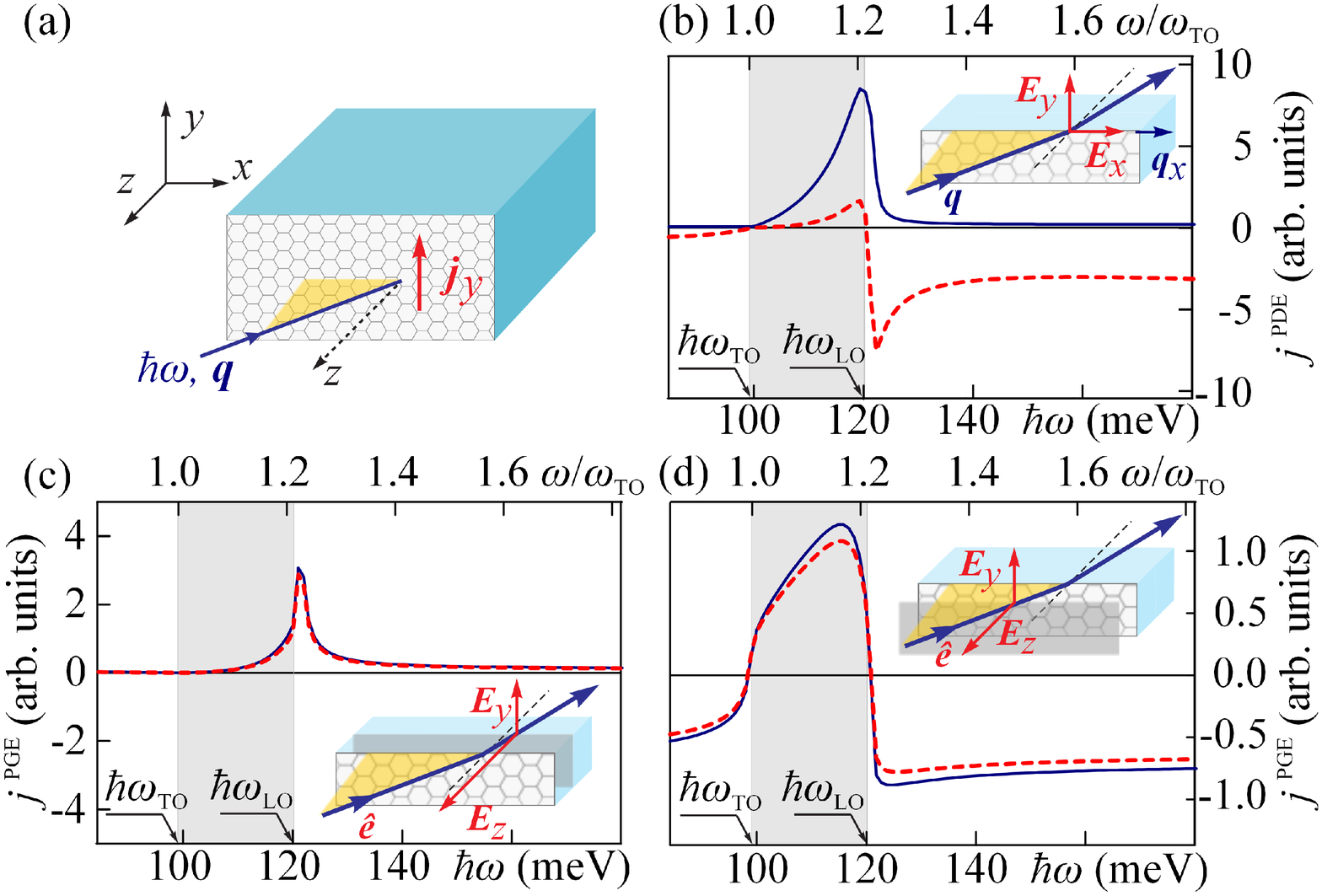}
\caption{(a) Geometry of the photocurrent generation. (b)-(d) 
Calculated spectral dependence of the linear (solid) and circular (dashed)  
photocurrents. (b) Photon drag effect. (c) and (d) Photogalvanic effect caused by
the electric field in the SiC side, $\bm E^{in}$, and in the air side of the air/SiC interface, $\bm E^{out}$, respectively.
The calculations are performed for $\theta_0=30^\circ$, $\gamma/\chi=0.9$, $T'/T=18$, the damping constant 
$\Gamma = 0.01 \, \omega_{TO}$ obtained from the reflection data, see Fig.~\ref{fig01_spectr}, 
and SiC high- and low-frequency dielectric constants $\varepsilon_\infty=6.52$, $\varepsilon_0=9.66$, respectively (see Ref.~\cite{Patrick}).    
The insets show components of the electric field and photon wave vector considered in 
the corresponding calculations.
} 
\label{fig04_theory}
\end{figure}

Figure~\ref{fig04_theory} (b) shows the calculated spectra of the
linear and circular photon drag effect. Calculations are carried 
out for the dominating contribution of the circular PDE - the fact which clearly follows from the data outside the reststrahl band. 
Within the reststrahl band the situation changes.
Due to the polarization transformation addressed above, 
the $T'$-contribution gives rise to the 
enhancement of the linear PDE. 
At the same time the decrease of the 
radiation helicity results in suppression 
of the circular photocurrent. 
While our calculations of the photon drag effect confirm the
enhancement/suppression of the linear/circular photocurrent within 
the reststrahl band they do not describe the double sign inversion and 
significant value of the linear photocurrent detected
outside the reststrahl band, see~Fig.~\ref{fig04_theory} (a).

\begin{figure}[t]
\includegraphics[width=\linewidth]{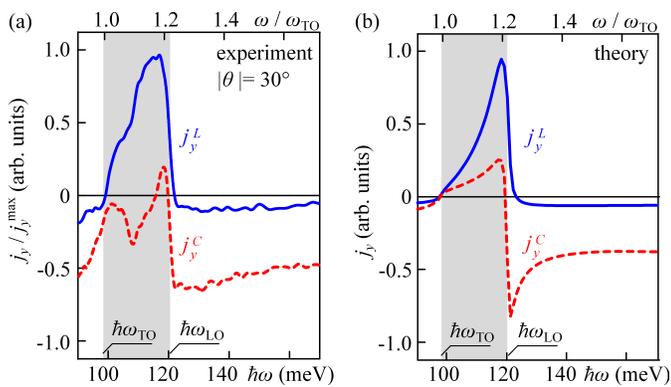}
\caption{Spectral behaviour of the linear (solid lines) and circular (dashed lines) photocurrents.
(a)  Experimental results. 
(b)  Calculated photocurrent for electric field components $E_x, E_y, E_z^{out}$, a ratio of PDE to PGE equal to $(\omega_{TO}/c)T'/\chi=0.3$ and the parameters given in the caption of Fig.~\ref{fig04_theory}.
%
%
}
\label{fig03_sum}
\end{figure}

To obtain a better agreement we consider the photogalvanic effect which, according to Ref.~\cite{PRB2010}, 
should yield a comparable contribution to the total 
photocurrent excited by infrared radiation.
It follows from  Eqs.~\eqref{j} that all PGE contributions require 
a normal component of the electric field, which is discontinuous 
at the interface. Therefore, from Eqs.~\eqref{j} and~\eqref{Ez} 
we obtain different solutions for the photocurrents
excited by the field in the air side of the air/SiC interface, $E_z^{out}$:
\begin{align}
\label{LPGE}
    j_y^{L} = -\sin{\theta_0} P_{\rm l} |E_0|^2 \left[ \chi \, {\rm Re}(t_p^* t_s)+ \gamma\, {\rm Im}(t_p^* t_s) \right] \Xi, \\
    j_y^{C} = -\sin{\theta_0} P_{\rm c} |E_0|^2 \left[ \gamma\, {\rm Re}(t_p^* t_s)+ \chi \, {\rm Im}(t_p^* t_s) \right] \Xi,
    \label{CPGE}
\end{align}
and from the field in the SiC side of the  interface, $E_z^{in}$:
\begin{align}
  \label{LPGEin}
        j_y^{L} = -\sin{\theta_0} P_{\rm l} |E_0|^2 \left[\chi \, {\rm Re}(t_p^* \varepsilon^* t_s)+ \gamma\, {\rm Im}(t_p^* \varepsilon^* t_s) \right] \Xi, \\
        j_y^{C} = -\sin{\theta_0} P_{\rm c} |E_0|^2 \left[\gamma \, {\rm Re}(t_p^* \varepsilon^* t_s)+ \chi \, {\rm Im}(t_p^* \varepsilon^* t_s) \right] \Xi.
          \label{CPGEin}
\end{align}
%
The resulting calculations, applying the same parameters as that used for calculations of PDE,
are shown in Figs.~\ref{fig04_theory}~(c),~(d).
It is seen that both linear and circular photogalvanic currents show a strong enhancement in the reststrahl band.
The solutions for the electric field in the substrate, see Fig.~\ref{fig04_theory}~(c), yield a sharp peak 
close to the longitudinal optical phonon energy only and, thus, do not describe the complex spectral behaviour 
of the photocurrent shown in Fig.~\ref{fig03_sum} (a). Moreover, we did not detect any sharp peak at LO frequency, 
which supports the conclusion that this contribution to the total photocurrent is negligible. 
A good agreement with the experiment is obtained for the linear photogalvanic current excited by 
the electric field in the air side of the air/SiC interface, see Fig.~\ref{fig04_theory}~(d).  
However, this solution does not describe the suppression of the circular photocurrent. 
We emphasize, that this disagreement can not be avoided simply by variation of the ratio between the coefficients 
$\gamma$ and $\chi$ describing the linear and circular photogalvanic currents, respectively.


Our calculations show that the overall reasonable 
agreement can only be achieved by considering a 
superposition of both PDE and PGE photocurrents. The corresponding results together with 
experimental data are shown in Fig.~\ref{fig03_sum}.
In fact, the calculations reflect all main features of the measured photocurrent, namely:
i)~Both circular and linear currents depend weakly on the frequency outside the reststrahl band;
ii)~The linear current changes its sign and has a broad peak within the whole reststrahl band;
iii)~The circular current is suppressed and has an asymmetric double-peak structure;
iv)~ The linear current dominates the circular one 
within the reststrahl band and vice versa in the outside.
We emphasize that, a satisfying agreement is obtained despite the fact that the 
calculations are carried out for the homogeneous functions $\varepsilon (\omega)$, abrupt interfaces 
and the influence of graphene on the radiation field was disregarded.
These simplifications used in our model or possible  $E_z$-electric field induced charge transfer between graphene and SiC~\cite{Kopylov} 
may be responsible for remaining small discrepancies, 
like a more pronounced dip in the circular current close to 
the center of reststrahl band.
A comparison of the photon drag and photogalvanic current components reveals that 
within the resonance the photon drag effect
strongly dominates in the total current whereas outside the resonance the main contribution comes from the linear photogalvanic effect. 
As for the circular photocurrent the PDE is responsible for the photoresponse in the whole studied spectral range.

To conclude, our results demonstrate that photocurrents in graphene deposited on a medium 
with a negative dielectric constant can be efficiently used for studies of how an electric field 
acts on the atomic scale. 
While the described approach is limited to the spectral  range defined by the reststrahl 
band of the substrate material, radiation of any desired frequency can be 
analyzed using negative dielectric constant of artificially made periodic structures, 
like metamaterials~\cite{metamaterials}.
%
Last but not at least, our findings demonstrate that 
optical and optoelectronic phenomena can be giantly enhanced  
in strictly two-dimensional systems and other nanoscale systems
if these structures are deposited on a substrate with a negative dielectric function.
This result is of importance for various kinds of applications in particularly 
those making use of effects proportional to higher orders of the electric field.

We thank S.A. Tarasenko and M.M. Voronov  for helpful discussions.
Support from DFG (SPP~1459 and GRK~1570),
Linkage Grant of IB of BMBF at DLR, RFBR, and POLAPHEN
is acknowledged.

\bibliographystyle{misha}
\bibliography{graphene}

\begin{thebibliography}{63}
\expandafter\ifx\csname natexlab\endcsname\relax\def\natexlab#1{#1}\fi
\expandafter\ifx\csname bibnamefont\endcsname\relax
  \def\bibnamefont#1{#1}\fi
\expandafter\ifx\csname bibfnamefont\endcsname\relax
  \def\bibfnamefont#1{#1}\fi
\expandafter\ifx\csname citenamefont\endcsname\relax
  \def\citenamefont#1{#1}\fi
\expandafter\ifx\csname url\endcsname\relax
  \def\url#1{\texttt{#1}}\fi
\expandafter\ifx\csname urlprefix\endcsname\relax\def\urlprefix{URL }\fi
\providecommand{\bibinfo}[2]{#2}
\providecommand{\eprint}[2][]{\url{#2}}





\bibitem{Novoselov2012} K.~S. Novoselov \textit {et al.}, 
Nature \textbf{490}, 192 (2012).


\bibitem{Dean2009} J.~J. Dean and H.~M. van Driel, 
Appl. Phys. Lett.  \textbf{95}, 261910 (2009).


\bibitem{Hendry2010} E. Hendry \textit {et al.}, 
Phys. Rev. Lett. \textbf{105}, 097401 (2010).


\bibitem{10.1063/1.3483872}
M.~Dragoman \textit {et al.}, 
Appl. Phys. Lett. \textbf{97}, 093101 (2010).


\bibitem{Hotopan11}
G.~Hotopan \textit {et al.}, 
Progress in Electromagneitc Research {\bf 118}, 57 (2011).


\bibitem{GlazovGanichev} 
M.~M. Glazov and S.~D.~Ganichev, arXiv:1306.2049v1 (2013).

\bibitem{Koppens2013}  K.~J. Tielrooij \textit {et al.}, 
Nature Phys. \textbf{9} , 248 (2013).


\bibitem{Bae2010} S. Bae \textit {et al.}, 
Nature Nanotech. \textbf{5}, 574 (2010).


\bibitem{GuT.:2012fk}
T.~Gu \textit {et al.}, 
Nature Photon., {\textbf 6}, 554 (2012).


\bibitem{Vicarelli2012} 
L. Vicarelli \textit {et al.}, 
Nature Mater. \textbf{11}, 865 (2012).


\bibitem{2011rangel} 
N. L. Rangel \textit {et al.}, 
J. Phys. Chem. C {\bf 115}, 12128 (2011). 


\bibitem{Chen2012} 
J. Chen \textit {et al.}, 
Nature \textbf{487}, 77 (2012).


\bibitem{Fei2012}
Z. Fei \textit {et al.}, 
Nature \textbf{487}, 82 (2012).


\bibitem{Jablan2009} 
M. Jablan, H. Buljan, and M. Soljacic, 
Phys. Rev. B \textbf{80}, 245435 (2009).


\bibitem{Koppens2011} 
F.~H.~L. Koppens, D.~E. Chang, and F.~J.~G. de Abajo, 
Nano Lett. \textbf{11}, 3370 (2011).


\bibitem{karch2010}
J.~Karch \textit {et al.}, 
Phys. Rev. Lett. \textbf{105}, 227402 (2010).


\bibitem{edge}
J.~Karch \textit {et al.}, 
Phys. Rev. Lett. \textbf{107}, 276601 (2011).


\bibitem{PRB2010}
C.~Jiang \textit {et al.}, 
Phys. Rev. B \textbf{84}, 125429 (2011).


\bibitem{Sun:2012ys}
D.~Sun \textit {et al.}, 
Phys. Rev. B {\bf 85}, 165427 (2012).


\bibitem{Prechtel:2012kx}
L.~Prechtel \textit {et al.}, 
Nature Comm. {\bf 3} 646, (2012).


\bibitem{cnt:book} 
A.~Jorio, G. Dresselhaus, M.~S. Dresselhaus, eds. in 
\textit{Advanced Topics in the Synthesis, Structure, Properties and Applications}, 
(Springer, 2008).

\bibitem{ivch_spi}
E.~L. Ivchenko and B.~Spivak,
Phys. Rev. B \textbf{66}, 155404 (2002).

\bibitem{PhysRevA.63.053808}
G.~Y. Slepyan \textit {et al.}, 
Phys. Rev. A \textbf{63}, 053808 (2001).

\bibitem{doi:10.1021/nl203003p}
G.~M. Mikheev \textit {et al.}, 
Nano Letters {\bf 12}, 77 (2012).

\bibitem{Zhang2011}
X.-L. Qi, S.\,C. Zhang, and X. L. Qi,
Rev. Mod. Phys. \textbf{83}, 1057 (2011).

\bibitem{Hosur2011}  
P. Hosur, 
Phys. Rev. B {\textbf 83}, 035309 (2011).

\bibitem{McIver2012} 
J.~W. McIver \textit {et al.}, 
Nature Nanotech. {\textbf 7}, 96 (2012).


\bibitem{Lupton} 
J. Lupton, 
Adv. Mater. \textbf{22}, 1689 (2010).


\bibitem{Cardona_book}
P. Y. Yu and M. Cardona, 
\textit{Fundamentals of Semiconductors: Physics and Materials Properties},  
(Springer, 4th ed., 2010).


\bibitem{Daas}  
B.~K. Daas \textit {et al.}, 
J. Appl. Phys. \textbf{110}, 113114 (2011).

\bibitem{Taminiau} 
T.~H. Taminiau \textit {et al.}, 
Nature Comm.  \textbf{3}, 979 (2012).


\bibitem{Borysiuk} 
J. Borysiuk \textit {et al.}, 
J. Appl. Phys. \textbf{105}, 023503 (2009).



\bibitem{naturenano} 
C. Drexler \textit {et al.}, 
Nature Nanotech. \textbf{8} 104 (2013).


\bibitem{Patrick} L. Patrick and W.J. Chouke, 
Phys. Rev. B \textbf{2}, 2255 (1970).

\bibitem{Ganichev02} 
S.\,D. Ganichev, E.\,L. Ivchenko, and W. Prettl,
Physica E {\bf 14}, 166 (2002).


\bibitem{footnote2}Note that the PGE current, being forbidden by symmetry in pristine graphene, is
allowed in graphene layer on a substrate due to broken $z \to -z$ symmetry (point group $C_{6v}$).
The samples under study are characterized by a strong structure inversion asymmetry as demonstrated in Ref.~\cite{naturenano}.  

\bibitem{footnote3} The polarization transformation is caused by existence of the imaginary parts 
of transmission coefficients $t_p$ and $t_s$ within the reststrahl band.


%





\bibitem{Kopylov} 
S. Kopylov, A. Tzalenchuk, S. Kubatkin, and V.I. Fal’ko,
Appl. Phys. Lett. \textbf{97}, 112109 (2010).

\bibitem{metamaterials} L. Solymar and  E. Shamonina, \textit{Waves in Metamaterials}  (Oxford Univ Press 2009).

\end{thebibliography}

\end{document}


\title{
Supplementary material for ''Probing of electromagnetic field on atomic scale by \\  photoelectric phenomena in graphene''
}
\author{
P.~Olbrich,$^1$ C. Drexler,$^1$ L.~E.~Golub,$^2$ S.~N.~Danilov,$^1$  V.~A.~Shalygin,$^3$ 
R.~Yakimova,$^4$ S.~Lara-Avila,$^5$ S.~Kubatkin,$^5$
B. Redlich$^6$, R. Huber,$^1$ S.~D.~Ganichev$^{1}$
}

\affiliation{$^1$ Terahertz Center, University of Regensburg,
93040 Regensburg, Germany}
\affiliation{$^2$ Ioffe Physical-Technical Institute, Russian Academy of Sciences,
194021 St.~Petersburg, Russia}
\affiliation{$^3$St.\,Petersburg State Polytechnic University, 195251 St.\,Petersburg, Russia}
\affiliation{$^4$
Link{\"o}ping University,
S-58183 Link{\"o}ping, Sweden}
\affiliation{$^5$
Chalmers University of Technology,
S-41296 G{\"o}teborg, Sweden}

\affiliation{$^6$ FOM Institute for Plasma Physics ``Rijnhuizen'', P.O.
Box 1207, NL-3430 BE Nieuwegein, The Netherlands}

\date{\today}

\maketitle

\section{Sample preparation}

Several large area graphene monolayer samples were grown on the Si-terminated face of a
4H-SiC(0001) semi-insulating substrate~\cite{supplLaraAvival09,supplKarch10} at $T = 2000$~$^\circ$C and 1~atm argon gas pressure~\cite{supplEmtsev09}.
The layers are $n$-doped due to the charge transfer from SiC with a measured
electron concentration in the range of $1.5 \times 10^{12}$ to
$7 \times 10^{12}$~cm$^{-2}$, Fermi energy $E_F$ ranging from~160
to 300~meV and mobility of about 10$^3$\,cm$^2$/Vs at room
temperature~\cite{supplKarch10,suppledge,supplJiang}.  
The samples are characterized 
by a strong structure inversion asymmetry
as demonstrated by the study of the magnetic quantum 
ratchet effect~\cite{supplnaturenano}.
Squares with dimensions of $5\times5$~mm$^2$ were patterned on graphene using
standard electron beam (e-beam) lithography and oxygen plasma
etching. Four metallic contacts on the periphery of graphene were
produced by straightforward deposition of Ti/Au (3/100~nm) through
a lithographically defined mask, followed by lift-off.
Ohmic contacts have been prepared at the center of the edges, with a resistance of about 2~k$\Omega$ between opposed contacts.

\section{Experimental}

The photocurrents were induced applying mid-infrared radiation of
the frequency tunable free electron laser ``FELIX'' at
FOM-Rijnhuizen in the Netherlands.~\cite{supplKnippels99p1578} The
laser operated in the spectral range between 7~$\mu$m and
15~$\mu$m (corresponding to photon energies from $\hbar \omega \approx
180$~meV to 90~meV). The output pulses of light from FELIX
were chosen to be $\approx$~2 ps long with peak power 
$P \approx 150$~kW, separated by 1~ns, in a train (or ``macropulse'') of
5~$\mu$s duration.  The beam has an almost almost Gaussian beam profile with a spot diameter 
of about 1~mm, which is measured by a pyroelectric 
camera~\cite{supplZiemann2000p3843}.
The laser spot was always smaller than 
the sample size allowing us to avoid illumination of contacts or 
sample edges and, consequently, 
to study only photocurrents generated in pristine graphene~\cite{suppledge}.
The radiation intensity $I$ and electric field $E$ on the sample during the 
micropulse were about 20~MW/cm$^2$ and 120~kV/cm, respectively. 
The macropulses had a repetition rate of 10~Hz. 
The room temperature photoresponse was studied in the directions 
perpendicular and parallel to the light incidence plane. 
The signals generated in the unbiased devices were measured via 
an amplifier with 20~MHz bandwidth
and recorded with a storage oscilloscope.
These measurements provide the full information about 
functional behaviour of the photocurrent, however,
the evaluation of the current magnitude
in response to such short pulses is not straightforward. 
Thus, to calibrate the 
photocurrent response we additionally measured  
the current  excited by radiation of a 
line-tunable \textit{cw} CO$_2$ laser with power $P$ of about 40~mW.
Though operating in a narrower spectral range (from 9.2 to 10.8~$\mu$m) 
it provides radiation in the vicinity of the upper limit of the reststrahl band 
and, therefore, is appropriated for the determination of 
the photocurrent value in the spectral region under study. 
The radiation power was controlled by a photon drag detector~\cite{supplGanichev84p20} and/or a mercury cadmium teluride detector.


\section{Reflection spectra}

\begin{figure}[t]
\includegraphics[width=0.99\linewidth]{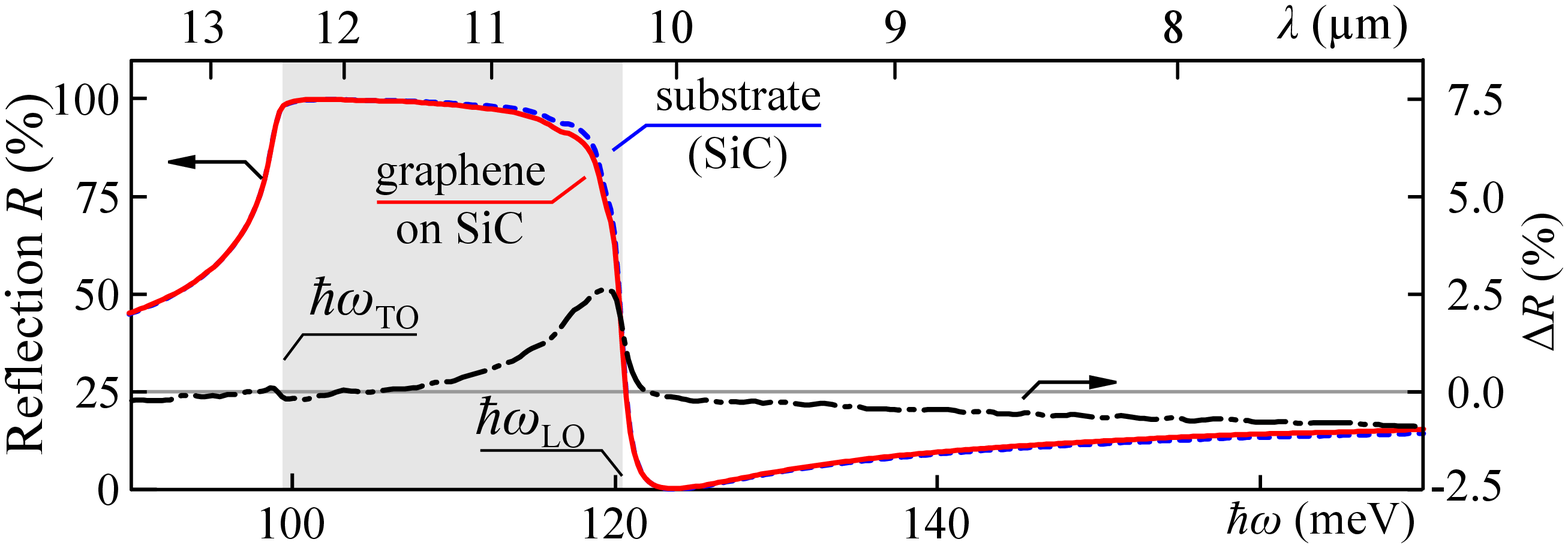}
\caption{
Dashed and solid lines show the reflection spectra of graphene on SiC substrate and pure SiC, 
respectively.  Dotted line shows the difference between the two curves.
The spectra are measured with Fourier spectrometer.
Grey area shows the range of the reststrahl band of SiC limited by 
the energies of transverse and longitudinal optical phonons.
} 
\label{figureS0}
\end{figure}

Figure~\ref{figureS0} shows the reflection spectra of graphene and SiC samples.
In both samples we observed almost total reflection $R(\omega)$ between the longitudinal, 
$\hbar \omega _{\rm LO}$,  and transversal, $\hbar \omega _{\rm TO}$, optical phonon energies of SiC. 
It is seen that reflection spectra of graphene layer on SiC slightly deviates from 
that of the pure SiC substrate (see doted line in Fig.~\ref{figureS0}). This result is 
in agreement with the data of Ref.~\cite{supplDaas} and is attributed to the substrate 
phonon-induced surface plasmon-polariton  formation in epitaxial graphene.

\section{Longitudinal photocurrent}

\begin{figure}[t]
\includegraphics[width=0.9\linewidth]{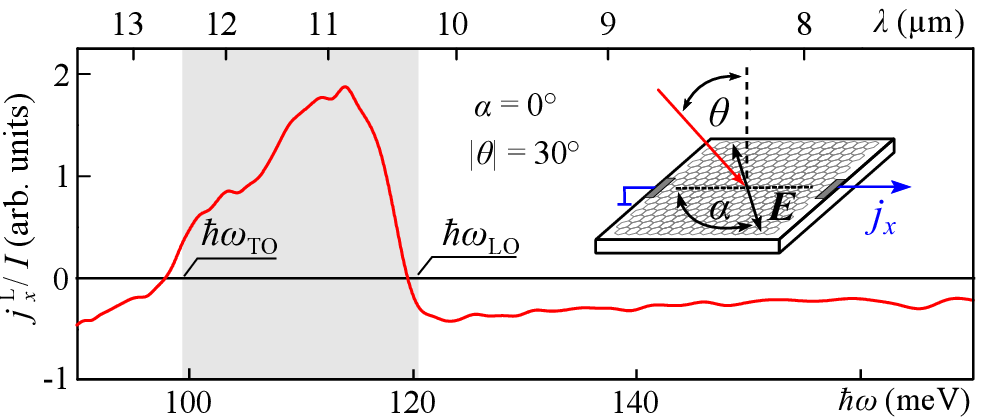}
\caption{Spectra of the longitudinal photocurrent $j_{x}^L$ measured 
for $|\theta| = 30^\circ$ and the azimuth angle $\alpha = 0^\circ$.
The data are obtained applying linearly 
polarized radiation of the  free electron laser ''FELIX''. 
Grey area indicates the range of the reststrahl band of SiC. 
The  inset shows the experimental geometry. 
} 
\label{figureS1}
\end{figure}

Besides the transversal photocurrent discussed in the paper, linearly 
polarized radiation also excites a current flowing along the light 
propagation direction, which varies as 
$j_x^L = L\cos{2\alpha}\sin{\theta} \, E_0^2$ (not shown). 
The spectral dependence of the latter photocurrent is shown in  
Fig.~\ref{figureS1} exhibiting resonance-like behaviour similar 
to the transversal one. For circularly polarized light the longitudinal  photocurrent is
not detected. The overall functional behaviour of the longitudinal 
photocurrent is in agreement with the phenomenological theory
and microscopic picture of the photon drag and photogalvanic effects~\cite{supplKarch10,supplJiang}. 
Calculations of the longitudinal photocurrent applying the same set of 
parameters as that used in the paper
are in a good agreement with the experiment.

\section{Fresnel transmission coefficients}

To obtain the frequency dependence of the required electric field components 
we use macroscopic Fresnel formulas. They are described by the corresponding Fresnel 
transmission coefficients which for oblique incident radiation are given by
\begin{equation}
    t_s = {2 \cos{\theta_0}\over \sqrt{\varepsilon-\sin^2{\theta_0}} + \cos{\theta_0}},
\end{equation}
\begin{equation}
    t_p = {2 \sqrt{\varepsilon} \cos{\theta_0}\over \sqrt{\varepsilon-\sin^2{\theta_0}} + \varepsilon\cos{\theta_0}}.
\end{equation}
Consequently, electric field components reflect the spectral behaviour of 
the dielectric function of the substrate which is given by~\cite{supplCardona_book}
\begin{equation}
\label{eps}
\varepsilon(\omega) = \varepsilon_\infty + {\varepsilon_0-\varepsilon_\infty \over 1 - (\omega/\omega_{TO})^2 - {\rm i}\omega\Gamma/\omega_{TO}^2}.
\end{equation}
At an oblique incidence of radiation on the dielectric media 
with $\varepsilon(\omega)$ the wavevector component in the surface plane is
$q_x=(\omega/c)\sin{\theta_0}$ is continuous while 
the normal wavevector component inside the 
medium $q_z^{in}=(\omega/c)(n + {\rm i}\varkappa)$ where~\cite{LL8}
\begin{equation}
    n = \sqrt{\sqrt{(\varepsilon' - \sin^2{\theta_0})^2 + {\varepsilon''}^2} + \varepsilon' - \sin^2{\theta_0} \over 2},
\end{equation}
\begin{equation}
\varkappa= \sqrt{ \sqrt{(\varepsilon' - \sin^2{\theta_0})^2 + {\varepsilon''}^2} -(\varepsilon' - \sin^2{\theta_0}) \over 2}.
\end{equation}
Here $\varepsilon'$ and $\varepsilon''$ are the real and imaginary parts of the 
dielectric function~\eqref{eps}, respectively.

\bibliographystyle{misha}
\bibliography{graphene}